\documentclass{aa}
\usepackage{psfig}
\usepackage{amsmath}

\newcommand{\kms}{km\thinspace s$^{-1}$}

\newcommand{\as}{$^{\prime\prime}$}
\def\to {$\rightarrow$}
\def\new#1 { #1 }
\def\cut#1 {}

\begin{document}

\title{High-excitation CO in a quasar host galaxy at $z=6.42$
\thanks{Based on observations obtained with the IRAM Plateau de Bure
Interferometer, and with the Effelsberg 100~m telescope. }}

\author{F. Bertoldi\inst{1},                
        P. Cox\inst{2}, 
        R. Neri\inst{3},
        C.L. Carilli\inst{4},
        F. Walter\inst{4},
        A. Omont\inst{5}, 
        A. Beelen\inst{2}, \\ 
        C. Henkel\inst{1},
        X. Fan\inst{6},
        Michael A. Strauss\inst{7},
        K.M. Menten\inst{1}
        }


\institute{Max-Planck-Institut f\"ur Radioastronomie, Auf dem H\"ugel 69,
           D-53121 Bonn, Germany 
      \and Institut d'Astrophysique Spatiale, 
	   Universit\'e de Paris XI, F-91405 Orsay, France
      \and IRAM,  300 rue de la Piscine, 38406 St-Martin-d'H\`eres, France  
      \and National Radio Astronomy Observatory, P.O. Box,
           Socorro, NM~87801, USA
      \and Institut d'Astrophysique de Paris, CNRS \& 
           Universit\'e Paris 6, 98bis bd. Arago, 75014 Paris, France     
      \and Steward Observatory, The University of Arizona, 
           Tucson AZ~85721, USA
      \and Princeton University Observatory, Princeton, NJ 08544, USA
}

\date{Received 30 May 2003 / Accepted July 2003}

\titlerunning{CO in a quasar host at $z=6.42$}
\authorrunning{F. Bertoldi et al.}

\abstract{We report the detection of high excitation CO emission from
the most distant quasar currently known, SDSS J114816.64+525150.3
(hereafter J1148+5251), at a redshift $z=6.419$.  The CO ($J=$6\to5)
and ($J=$7\to6) lines were detected using the
IRAM Plateau de Bure interferometer, showing a width
of $\approx 280 \, \rm km \, s^{-1}$.  An upper flux limit for
the CO ($J=$1\to0) line was obtained from observations with the
Effelsberg 100-meter telescope.  Assuming no gravitational
magnification, we estimate a molecular gas mass of $\approx 
2 \times 10^{10} \, M_{\odot}$. 
Using the CO (3\to2) 
observations by Walter et al.~(2003), a comparison of the
line flux ratios with predictions from a large velocity gradient model
suggests that the gas is likely of high excitation, at densities
$\sim 10^5 \rm ~cm^{-3}$ and a temperature $\sim 100$~K.  
Since in this case 
the CO lines appear to have moderate optical depths, the gas 
must be extended \new{over a few kpc.}
The gas mass detected in J1148+5251 can fuel star formation at the
rate implied by the far-infrared luminosity for less than 10 million years, 
a time comparable to the dynamical time of the region. The gas must
therefore be replenished quickly, and metal and dust enrichment must
occur fast.
The strong dust emission and massive, dense gas reservoir at $z\sim
6.4$ provide further evidence that vigorous star formation is co-eval
with the rapid growth of massive black holes at these early epochs of
the Universe.
}
\maketitle
\sloppy

\section{Introduction}
\label{sec:Introduction}

The luminous quasars at redshifts $z>6$ found in the Sloan Digital Sky
Survey by Fan et al.\ (2001, 2003) provide a unique opportunity
to study the formation of massive objects during the epoch at which
the intergalactic medium was being reionized by the first luminous
sources (Becker et al.~2001; Kogut et al.~2003; Cen 2003).  Studying
signatures of star formation in these exceptional objects is also of
great interest to test whether the correlation between the central
black hole mass and the stellar bulge mass observed in local spheroids
(Magorrian et al.~1998; Gebhardt et al.~ 2000) can be traced to the
early formation stages of quasars and their host galaxies.

J1148+5251, at a redshift of $z=6.42$ (Fan et al.\ 2003), is the most
distant quasar known, observed only $\approx 850$ million years after
the Big Bang (we adopt 
$H_0=71\rm~km~s^{-1}~Mpc^{-1}$, $\Omega_\Lambda=0.73$ and
$\Omega_m=0.27$ -- Spergel et al.~2003).  Optical, radio and
millimeter observations indicate that J1148+5251 could be weakly
amplified by an intervening lens (Fan et al.~2003; White et al.~2003;
Bertoldi et al.~2003), but in what follows, we will assume no
lens amplification.  J1148+5251 is an very luminous quasar
($M_{1450}=-27.8$, $L_{\rm bol} \sim 10^{14} \, L_\odot$) powered by a
supermassive ($\approx 3 \times 10^9 \, M_\odot$) black hole radiating
close to its Eddington luminosity (Willott et al.\ 2003). If the
mass of the dark matter halo associated with J1148+5251 is
proportional to the black hole mass in a way similar to what is found
in local spheroids (Shields et al.~2003), its mass would be $\approx 2
\times 10^{12} \, M_\odot$,
and J1148+5251 would be among the most massive
collapsed structures to have formed in the early Universe (e.g.,
Haiman \& Loeb 2001).

The recent detection of thermal dust emission in J1148+5251 (Bertoldi
et al.~2003) implies a far-infrared luminosity of $\approx 10^{13} \,
L_\odot$ and a dust mass of $\approx 7 \times 10^8 \, M_\odot$.
If the dominant heating mechanism is radiation from young stars, then the
star formation rate implied from the FIR luminosity is $\sim 3000 \,
M_\odot \, \rm yr^{-1}$ \cut{(see \S3.4 of Omont et al.~2001)}, which
requires vast amounts of molecular gas to be maintained.

Molecular gas \cut{masses} in excess of $10^{10} \, M_\odot$ was detected
through their CO emission in fifteen $z>2$ far-infrared ultraluminous
($L_{\rm FIR}> 10^{12} \, L_\odot$) radio galaxies and quasars (e.g., Cox, Omont,
\& Bertoldi~2002). At $z>4$, CO emission was detected towards four quasars
(Omont et al.~1996; Ohta et al.~1996; Guilloteau et al.~1997, 1999;
Cox et al.~2002).  The CO emission was resolved in BR~1202$-$0725
(Carilli et al.~2002) at $z=4.69$, the highest
redshift CO detected so far, and PSS~2322$+$1944 at $z=4.12$ (Carilli et
al.~2003). The extended nature of the CO provides the most direct
evidence for active star formation in the host galaxies of distant
quasars, and indicates that black hole accretion and star-formation are
closely related.

To explore the growth of massive black holes and their associated
stellar populations at the end of the ``dark ages'', we have searched
for CO emission toward J1148+5251.  We here report the detection of CO
(6\to5) and (7\to6) line emission.
In a separate study, Walter et al.~(2003) report the discovery of CO
(3\to2) emission using the Very Large Array (VLA).
\cut{we use those results below.}

\section{Observations}
\label{sec:observations}

\begin{figure}[tb]
 \centerline{\psfig{figure={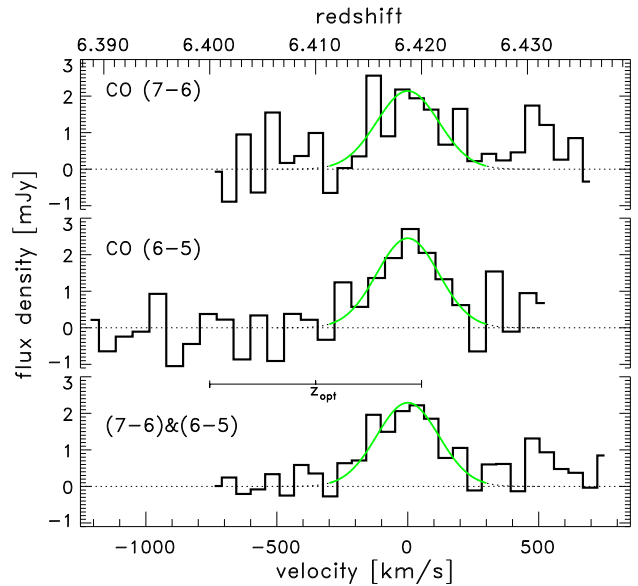},width=8.5cm}}   
  \caption{ J1148+5251 spectra of CO (6\to5), (7\to6), \new{and their average, 
  binned to 64, 55, and 55 \kms, respectively, four times the original
  spectral resolution. Zero velocity corresponds 
  to the centroid of the (6\to5) line at 93.206~GHz. 
  Gaussian fits with FWHM$=279$\kms\ are shown as light lines.}
\cut{The redshift derived by Willott et al.~(2003) from the Mg{\sc ii}
$\lambda 2799$ emission line is indicated.}
}
\label{fig:spectra} 
\end{figure}

\begin{figure}[tb]
 \centerline{\psfig{figure={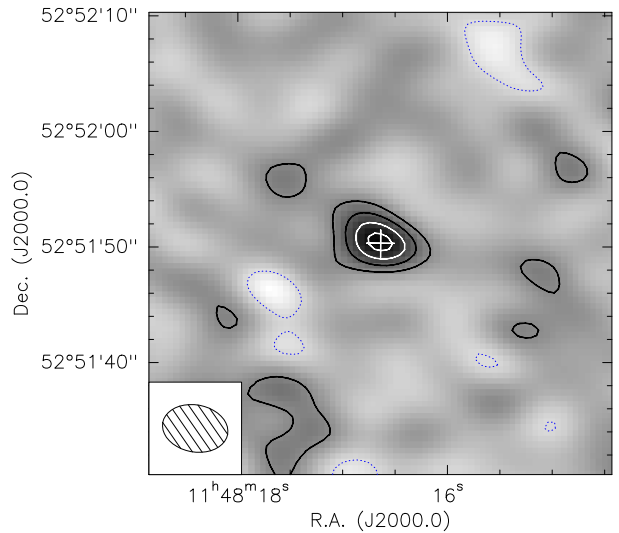},width=8.cm}}   
  \caption{Velocity-integrated ($\pm 220$\kms) map of the averaged CO (6\to5)
   and (7\to6) emission.
   Contour steps are 0.34~mJy/beam $=2\sigma$. 
   The cross indicates the optical position. 
   }
\label{fig:map} 
\end{figure}

Observations of the CO (7\to6) and (6\to5)
emission lines were made with the IRAM Plateau de Bure interferometer
between March and May 2003. We used the 6 antenna D configuration
which results in a beam of $5.7^{\prime\prime} \times
4.1^{\prime\prime}$ at 3.2~mm.  The 3~mm receivers were tuned in
single sideband and the typical SSB system temperatures were $\approx
150$~K.
The total integration time was 14 hours for CO (6\to5), and 22
hours for CO (7\to6). 
\cut{The data were reduced, calibrated and analyzed
using the standard IRAM programs C{\small LIC} and M{\small APPING}.}
The spectra are displayed in Fig.~\ref{fig:spectra}, and the
image of the averaged data
is shown in Fig.~\ref{fig:map}.
Within the astrometric uncertainties
of $\pm 0\farcs3$, the CO emission coincides with the optical position
given by Fan et al.~(2003). At the 5\as\
resolution of our Plateau de Bure observations, the CO emission is
unresolved, consistent with the VLA CO(3\to2) detection, which is
unresolved at $1\farcs5$ (Walter et al 2003).

\begin{table*}[ht]
\caption{Properties of the CO lines observed toward SDSS J1148+5251}
\begin{center}
\begin{tabular}{ccccccccc}
\hline
\hline 
Line &  $\rm \nu_{\rm rest}$ & $\rm \nu_{\rm obs}$ & $z_{\rm CO}$ & peak int.  &  $\Delta v_{\rm FWHM}$   &  $I_{\rm CO}$  & $L^\prime_{\rm CO}$ & $L_{\rm CO}$  \\

     & \multicolumn{2}{c}{[GHz]} &  & [mJy] & [$\rm km \, s^{-1}$] &  [Jy~km~s$^{-1}$] & [$\rm 10^{10} \, K \, km \, s^{-1} \, pc^2$] & [$10^8 \, L_\odot$] \\ 

\hline \\[-0.2cm]

CO (7\to6)& 806.651 & 108.729 & 6.4192$\pm$0.0009 & 2.14 & 279$^\dagger$ & 0.64$\pm$0.088 & 1.73$\pm$0.24 & 2.92$\pm$0.40 \\
CO (6\to5)& 691.473 & 93.204  & 6.4187$\pm$0.0006 & 2.45 & 279~  & 0.73$\pm$0.076 & 2.69$\pm$0.24 & 2.86$\pm$0.25 \\
CO (3\to2)& 345.795 & 46.610  & 6.419$\pm$0.004   & 0.6  & 320$^\ddagger$ & 0.18$\pm$0.02 & 2.68$\pm$0.27 & 0.35$\pm$0.04 \\
CO (1\to0)& 115.271 & 15.537  & --                & $<0.36$ &  --  & $<0.11^\dagger$ & $<14.2$ & $<0.070$ \\
\hline
\end{tabular}
\end{center}
NOTE. -- For J1148+5251, 
the apparent CO line luminosity is given by 
$L^\prime_{\rm CO} = 3.2 \times 10^4 I_{\rm CO} \nu_{\rm obs}^{-2}$, 
the intrinsic line luminosity
$L_{\rm CO}=4.2\times 10^{6}  I_{\rm CO} \nu_{\rm obs}$, 
in the units given above (see Solomon et al.~1997). Upper limits are 3$\sigma$.
$^\dagger$ Adopting the line width of CO (6\to5).
$^\ddagger$ Line width corresponds to the 50~MHz channel width of the VLA 46.6~GHz observations. 
\label{table:properties}
\end{table*}

The CO (6\to5) and (7\to6) lines are detected at centroid frequencies
\cut{of 93.206~GHz and 108.724~GHz,} corresponding to a redshift 
$6.419$ \cut{for CO(6\to5)} (Table~\ref{table:properties}).
Within the uncertainties this agrees with the $z=6.41\pm0.01$ of the
Mg{\sc ii} $\lambda 2799$ line (Willott et al.~2003)
(Fig.~\ref{fig:spectra}).  The CO redshift, which is likely to
correspond to the systemic redshift of the quasar, differs
significantly from the range $z=6.36-6.39$ derived from high
ionization UV lines (White et al.~2003), which trace high velocity
($\rm \ge 1000 \, km \, s^{-1}$), blue-shifted gas related to the
quasar activity. 

\new{No continuum emission was detected in our coadded 3~mm data,
which \cut{also} includes observations at other frequencies, but excludes
the continuum redward of the CO lines, where
weak line emission may be present. At the position of J1148+5251 we
obtain a continuum flux of $0.09\pm 0.13$ mJy.
At 43~GHz the continuum remains undetected with $-31 \pm
57~\mu$Jy.  These upper limits
are consistent with the measured 250~GHz flux density of $\rm 5.0\pm0.6 \,
mJy$ (Bertoldi et al.~2003), if we adopt a grey body spectrum with
temperature $>50$~K and dust emissivity $\propto \nu^{2}$.}

\new{We fit Gaussians to the line spectra within
$\pm 300$ \kms\ of the centroid, with no baseline subtraction.  
The best fit line widths of the three spectra shown in Fig. 1
range between 280 and 320 $\rm km \, s^{-1}$, similar to the
widths found in other high redshift quasars (e.g., Cox et al.~2002).
The width and centroid of the best Gaussian fit to the (6\to5) line were
adopted for the Gaussian fit to the lower quality (7\to6) line to
determine its flux.}  \cut{A free fit would yield a somewhat broader line
with a 10\% larger flux,
and an integration within $\pm 250$ \kms\ yields a flux within 1\% of
that from the former Gauss fit.} \new{The CO (6\to5) and (7\to6) line fluxes
are determined at strong confidence levels of 10$\sigma$ and 7$\sigma$,
respectively.}

We searched for  CO (1\to0) emission 
using the Effelsberg 100-meter telescope in March and April 2003
with a 1.9\,cm HEMT receiver ($T_{\rm sys}\sim 40$~K on a $T_A^*$
scale, aperture efficiency $\sim$40\%, beam width 60\arcsec, position
switching mode).
The integration time was $\sim$50 hours, yielding a r.m.s. of $T_A^*
\sim 0.4$~mK ($\sim 0.4$ mJy per 24 km s$^{-1}$).  No CO (1\to0)
emission was detected at the redshift found for the higher CO
transitions (Table~\ref{table:properties}).

\section{Discussion}
\label{sec:discussion}

To constrain the physical conditions of the molecular gas in
J1148+5251 we compared the observed CO line flux ratios with those
predicted by a one-component large velocity gradient (LVG) model (Mao
et al.~2000). The line flux ratios are determined by
the gas density, temperature, and the  optical depth in the CO
lines, i.e., the column density of CO per velocity
interval.
\new{The large flux ratio between the
(6\to5) and (3\to2) lines implies that the gas has a high excitation.
The lower excitation of the $J=7$ level 
suggests a moderate optical depth (Fig.~3).}

\begin{figure}[tb]
\centerline{\psfig{figure={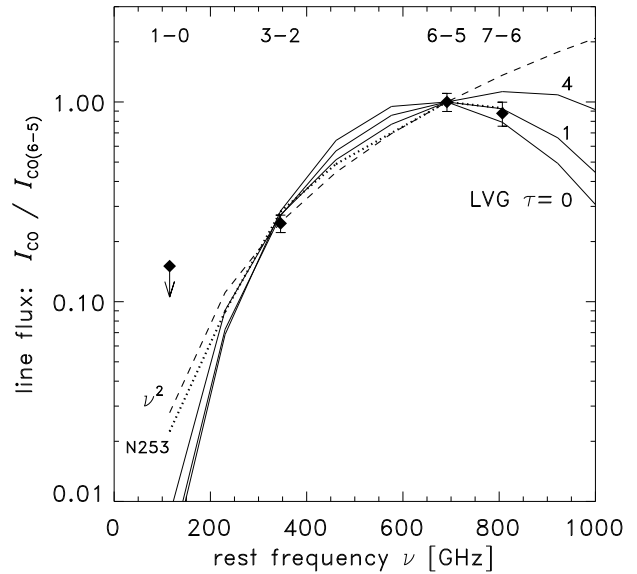},width=8.5cm}}   
  \caption{Integrated line flux, $I_{\rm CO}$, \new{normalized to CO
  (6\to5).} Diamonds show the values for J1148+5251. The dashed line
  shows line flux increasing as $\nu^2$, which is expected for
  optically thick conditions. \new{The solid lines show LVG models with
  $T_{\rm kin} = 120 \, \rm K$ and $n({\rm H_2}) = 4.5 \times 10^4 \,
  \rm cm^{-3}$,} with different maximum optical depth in the CO
  lines.
  The dotted line shows the line flux distribution observed for
  the starburst nucleus of NGC~253 (Bradford et al.~2003).
   }
\label{fig:excitation} 
\end{figure} 

High gas densities are typical of the molecular gas present
in the nuclear regions of nearby starburst galaxies (Solomon \& Downes
1998).
The most extreme conditions so far were found in the starburst
nucleus of NGC~253 (Fig.~3), where the CO excitation
is similar to that in J1148+5251.
Detailed LVG modeling by Bradford
et al.~(2003) indicate that the CO, $\rm ^{13}CO$, and H$_2$
data of NGC~253 are consistent
with $n(\rm H_2) = 4.5\times 10^4\rm cm^{-3}$ and $T=120$~K, and CO
line optical depths $\tau<4$ (Fig.~\ref{fig:excitation}). 
\new{With only three line fluxes and one upper limit, we cannot constrain the
physical conditions of the gas in J1148+5251 as tightly: the data can be
fit both with low-opacity models, in which temperature and density are
degenerate,  $T n^{1/2} \approx 2.5\times 10^4 \rm ~K~cm^{-1.5}$,
and with high-opacity lower-excitation models.
With the high excitation temperatures for the gas in NGC~253 and J1148+5251
the cosmic background temperature (3~K and 20~K, respectively) does not affect the gas 
excitation notably.} 


To infer the total molecular mass from the CO emission in the absence
of constraints on the CO abundance, one typically adopts an empirical
conversion factor, $\alpha$, between the apparent CO (1\to0) line 
luminosity $L'_{\rm CO}$,
and the total molecular mass, $M_{\rm H_2}$. 
\new{However, this conversion may depend on the CO excitation.}
\cut{However, the extent to which CO (1\to0) is a good measure of $M_{\rm H_2}$ 
represents the total molecular gas mass much depends on the CO excitation,
making such estimates inaccurate.}
In nearby starbursts with moderate gas
excitation, Downes \& Solomon~(1998) derive
$\alpha=0.8 \, M_\odot \rm \, (K \, km \, s^{-1} \, pc^2)^{-1}$.  The
high excitation and moderate line opacities of our one-component
LVG model for J1148+5251
predict a CO (1\to0) line flux much lower than that of an
optically thick distribution (Fig.~3). 
\new{As observed for NGC~253, it is likely that an optically thick, low-excitation 
molecular component adds to the lower $J$ level populations.
Given this uncertainty,} and since a value of $\alpha$
for very high excitation conditions is unknown, we do not estimate the mass
on the \new{one component} LVG predictions, but extrapolate to 
$L'_{\rm CO (1\rightarrow0)} = 2.7 \times
10^{10} \, \rm K \, km \, s^{-1} \, pc^{2}$, with the assumption of a
constant line brightness temperature (the optically thick case) from
$J=1$ to 6. With the quoted conversion factor, we find 
$M_{\rm H_2} \approx 2 \times 10^{10} \, M_\odot$.
\cut{in agreement with the Effelsberg CO(1\to0) upper flux limit which yields
$M_{\rm H_2} < 10^{11} \, M_\odot$.}
We then estimate the gas to dust mass ratio $M_{\rm H_2}/M_{\rm dust} \approx
30$, which is similar to the values found for local ULIRGs and
other high redshift quasars (e.g., Guilloteau et al.~1999; Cox et
al.~2002).


The \new{minimum area} of the molecular region can be estimated from the ratio of the
observed line brightness temperature (11 mK for CO (6\to5)) and the
intrinsic line brightness, which is 23 and
56~K in the LVG models with $\tau=4$ and 1, respectively
(Fig.~\ref{fig:excitation}).  With a 5\as\ beam the corresponding
source radius (assuming uniform coverage) is 0.1--0.15\as, or
560--840~pc. Placing $2\times 10^{10}M_\odot$ in a volume of radius
560~pc gives an average CO column density $1.2\times 10^{20}(X_{\rm
CO}/10^{-4})\rm cm^{-2}$, where $X_{\rm CO}$ is the CO abundance
relative to H$_2$.
If the observed \cut{high-$J$} CO lines have moderate
optical depth, $\tau<4$, the CO column is $<2\times 10^{19}\rm
cm^{-2}$, which would either require a low CO abundance, $X_{\rm
CO}<2\times 10^{-5}$, or a larger volume. Considering the low gas-to-dust
ratio estimated above and the high metalicities implied by the
optical lines (Fan et al.~2003), a CO abundance much lower than
the Galactic $\sim 10^{-4}$ seems unlikely. Rather, a larger radius
of $\sim 1400$ pc for the gas distribution could account for the
moderate CO line opacities.

If the molecular gas forms an inclined disk (angle $i$
relative to the sky plane) in Keplerian rotation about a spherical
mass, the line width
and a minimum source radius between 560 and 1400~pc yield a minimum
gravitational mass enclosed by the disk of $(2-6) \times 10^9
\sin^{-2} i \, M_\odot$. \new{For large inclination angles this mass would
not be} much larger than that of the
black hole, and a factor 4--10 smaller than the gas mass implied by the
line intensities.  The latter may have been overestimated given the
approximate nature of our estimate; alternatively, the CO disk inclination is
close to the sky plane, $i\sim 20-30$ deg, which is more likely
considering the large dust mass, and the fact that the AGN is optically
unobscured.


The detection of large amounts of dense molecular gas in J1148+5251
supports the conjecture that the strong far-infrared luminosity seen
from many quasars arises from extended star forming regions, and is
not due to heating from the AGN (Omont et al.~2001, 2003; Carilli et
al.~2001).
Although for J1148+5251 the emission remains spatially unresolved, the
large masses of warm CO are unlikely to be heated by the AGN at 
a kpc distance.  With the estimated mass of molecular gas of $\sim 2
\times 10^{10} \, M_\odot$ star formation in J1148+5251 could be
sustained at the \new{rate $\sim 3000 \, M_\odot \, \rm yr^{-1}$ implied} by
the far-infrared luminosity for a short time only, \new{ $<10$ million}
years. This is comparable to the estimated duty cycle time of quasars
(e.g. Wyithe \& Loeb 2003), and to the dynamical time of the star
forming region, which implies a rapid gas depletion unless the system
continues to accrete gas at a high rate.
If the replenishing gas is of low-metalicity, the short depletion time
suggests that the enrichment with heavy elements and dust is rapid,
which leaves only supernovae and winds from the most 
massive stars as possible sources. 

Our low estimate for the dynamical to luminous mass ratio excludes the
presence of a large stellar mass within the volume of the CO emission.
The duration of star formation at the present rate could therefore not
have been much longer than $10^7$ yr, unless the starburst does not
form many long-lived, low-mass stars, in which case the star formation
rate would have been overestimated and the depletion time could be
longer.

The dynamical mass is an order of magnitude smaller than the bulge
mass deduced from the correlation between the black hole mass and the
bulge mass (or velocity dispersion) in local spheroids (Magorrian et
al.~1998). This could be due to a biased selection of a
non-representative, bright quasar, or it confirms a tendency for the
stellar to black hole mass ratio to decrease at higher redshifts (Rix
et al.~1999), possibly due to self-regulating star formation
mechanisms (Wyithe \& Loeb 2003).

\acknowledgements 

We thank the IRAM Plateau de Bure staff and the
Effelsberg operators for their great support in the observations,
and F.~Combes for helpful comments.

\end{document}